\documentclass[12pt,fleqn, a4paper]{article}

\newcommand{\ra}{\rightarrow}

\newcommand{\bra}{\langle} \newcommand{\ket}{\rangle}
\newcommand{\be}{\begin{equation}}
\newcommand{\ee}{\end{equation}}
\newcommand{\bea}{\begin{eqnarray}}
\newcommand{\eea}{\end{eqnarray}}

\newcommand{\E}{\mbox{e}}

\newcommand{\ep}{\qquad {\vrule height 10pt width 8pt depth 0pt}}

\newcommand{\grintl}{[\kern-.18em [}
\newcommand{\grintr}{]\kern-.18em ]}

\newtheorem{lem}{Lemma}[section]
\newtheorem{prop}{Proposition}[section]
\newtheorem{thm}{Theorem}[section]

\usepackage{amsfonts}
\def\R{{\mathbb R}}

\def\0{{\mathbb O}}
\def\un{{\mathbb I}}
\def\T{{\mathbb T}}
\def\Z{{\mathbb Z}}
\def\N{{\mathbb N}}
\def\C{{\mathbb C}}
\def\E{{\mathbb E}}
\def\P{{\mathbb P}}

\def\I{{\mathcal I}}

\begin{document}

\title{Localization for Random Unitary Operators}
\author{Eman Hamza\\
University of Alabama at Birmingham\\
Department of Mathematics
 CH 452\\
Birmingham, Al 35294-1170\\ U.S.A.
\and
Alain Joye\\
Institut Fourier
\\ Universit\'e de Grenoble, BP 74
\\38402 Saint-Martin d'H\`eres\\
France\\
 \and
G\"unter Stolz\footnote{partially supported through US-NSF grant DMS-0245210}\\
University of Alabama at Birmingham\\
Department of Mathematics
 CH 452\\
Birmingham, Al 35294-1170\\ U.S.A.
}
\date{  }
\maketitle \abstract{We consider unitary analogs of
$1-$dimensional Anderson models on $l^2(\Z)$ defined by the
product $U_\omega=D_\omega S$ where $S$ is a deterministic unitary
and $D_\omega$ is a diagonal matrix of i.i.d.\ random phases. The
operator $S$ is an absolutely continuous band matrix which depends
on a parameter controlling the size of its off-diagonal elements.
We prove that the spectrum of $U_\omega$ is pure point almost
surely for all values of the parameter of $S$. We provide similar
results for unitary operators defined on $l^2(\N)$ together with
an application to orthogonal polynomials on the unit circle. We
get almost sure localization for polynomials characterized by
Verblunski coefficients of constant modulus and correlated random
phases. }


\setcounter{equation}{0}
\section{Introduction}

Unitary operators displaying a band structure with respect to a
distinguished basis of $l^2(\N)$ or $l^2(\Z)$ appear in different
contexts. For example, such operators describe the quantum
dynamics of certain models in solid state physics, see e.g.
\cite{bb}, \cite{bhj} and references therein. Unitary band
matrices also appear naturally in the study of orthogonal
polynomials on the unit circle with respect to a measure, as was
recognized recently in \cite{cmv}. For a detailed account on
orthogonal polynomials on the unit circle, see \cite{s} (which is
briefly surveyed in \cite{s3}). In both situations, the spectral
properties of these unitary infinite matrices play an important
role.

The spectral analysis of a certain set of deterministic and random
unitary operators with band structure is undertaken in \cite{bhj}
and \cite{j}, \cite{j2}. The random cases studied in the first two
papers concern a set of matrices on $l^2(\Z)$ which are (up to
unitary equivalence) of the following form: $U_\omega=D_\omega S$
where $S$ is a deterministic unitary and $D_\omega$ is a diagonal
matrix of random phases, diag$\{e^{-i\theta_k^\omega}\}$. The
operator $S$ is an absolutely continuous band matrix which depends
on a parameter $t\in ]0,1[$ which controls the size of its
off-diagonal elements, see Section \ref{mod} below.

When the phases are i.i.d random variables, typical results
obtained for discrete one-dimensional random Schr\"odinger
operators are shown in \cite{bhj} and \cite{j} to hold in the
unitary setting as well. For instance, the availability of a
transfer matrix formalism to express generalized eigenvectors
allows to introduce a Lyapunov exponent, to prove a unitary
version of the Ishii-Pastur Theorem, and get absence of absolutely
continuous spectrum \cite{bhj}. A density of states can be
introduced and a Thouless formula is proven in \cite{j}. Since
these operators can be naturally considered as unitary analogs of
the self-adjoint Anderson model, generalizations to $d$-dimensions
are introduced in \cite{j2}. Their localization properties are
studied by means of an adaptation to the unitary setup of the
fractional moment method due to Aizenman and Molchanov \cite{am}.
In \cite{j2} it is shown for arbitrary dimension that localization
takes place if the common distribution of the phases is absolutely
continuous and the parameters of the $d$-dimensional deterministic
unitary $S$ are such that $S$ is close to the identity. This is
the unitary analog of the familiar large disorder regime under
which localization holds for the $d$-dimensional Anderson model.
When applied to the one-dimensional case, this yields localization
only if the parameter $t$ is sufficiently small.

One of the goals of the present paper is to prove localization for
all values of the parameter $t$, thereby completing the analogy
with the self-adjoint one dimensional Anderson model, where
localization holds without an additional disorder assumption. We
also complete the picture by considering products of the same sort
on $l^2(\N)$, for which we prove localization as well.

Another motivation comes from the application of such results to
orthogonal polynomials on the unit circle (OPUC), with respect to
a measure $d\mu$. These polynomials are characterized by a
sequence of complex numbers $\{\alpha_k\}_{k\in\N}$,  such that
$|\alpha_k|<1$, $k\in\N$, called the Verblunski coefficients.
These coefficients allow to construct a unitary infinite matrix on
$l^2(\N)$, the so-called CMV matrix \cite{cmv}, which is the
equivalent in the OPUC setting of the Jacobi matrix for orthogonal
polynomials on the real line. This matrix represents
multiplication by $z$, $z\in S^1$, on $L^2(S^1, d\mu)$ so that its
spectral measure is $d\mu$, see \cite{s}. When the Verblunski
coefficients $\{\alpha_k(\omega)\}_{k\in\N}$ are random, the fine
structure of the corresponding random measure $d\mu_\omega$ is of
interest. As mentioned in \cite{j}, when $|\alpha_k(\omega)|=r$
for all $k\in\N$, and only the phases of $\alpha_k(\omega)$ are
random, the CMV matrix is unitarily equivalent to the product
$D_\omega S$ on $l^2(\N)$ which, modulo boundary conditions at
site $0$, is of the form considered  on $l^2(\Z)$ above. The point
now is that the random phases of the coefficients
$\alpha_k(\omega)$ are correlated if the phases of the diagonal
matrix $D_\omega$ are independent.

Therefore we get as a corollary of our general analysis that
localization takes place for random OPUC with certain types of
correlated Verblunski coefficients. Let us mention here that
previous localization results for CMV-matrices provided in
\cite{gt}, \cite{t}, \cite{s}, \cite{ps}, and \cite{su}
essentially consider independent Verblunski coefficients.

\section{The Model and Main Result}\label{mod}\setcounter{equation}{0}

We introduce in this section the set of infinite random unitary matrices
on $l^2(\Z)$ we will be interested in. We focus on the properties
related to the localization result we are to prove.
For further details and generalizations, we refer the reader to
\cite{bhj}, \cite{j} and \cite{j2}.

Let us denote by $|k\ket$ the unit vector at site $k\in \Z$, so
that $\{|k\ket\}_{k\in\Z}$ forms an orthonormal basis of $l^2(\Z)$.
We introduce a probability space $(\Omega, {\cal F}, \P)$, where
$\Omega$ is identified with $\{{\T}^{\Z} \}$, $\T = \R / 2\pi\Z$
being the torus, and $\P=\otimes_{k\in\Z}\P_k$, where $\P_{k}=\nu$
for any $k\in\Z$ and $\nu$ is a fixed probability measure on $\T$,
and ${\cal F}$ the $\sigma$-algebra generated by the cylinders. We
introduce a set of random variables on $(\Omega, {\cal F}, \P)$  by
\bea\label{beta} \theta_k: \Omega \rightarrow \T,
  \ \ \mbox{s.t.} \ \ \theta_k^\omega=\omega_{k}, \ \ \ k\in \Z.
\eea These random variables $\{\theta_k\}_{k\in\Z}$ are thus i.i.d.\
on $\T$.

We consider
unitary operators of the form
\be\label{1d}
 U_{\omega}=D_{\omega}S, \,\,\,\mbox{ with }
D_{\omega}=\mbox{ diag }\{e^{-i\theta_k^\omega}\}
\ee
and
\be\label{s0}
S=\pmatrix{\ddots & rt & -t^2& & & \cr
              & r^2& -rt  & & & \cr
              & rt & r^2 & rt & -t^2& \cr
              & -t^2 &-tr & r^2& -rt& \cr
              & & & rt &r^2 & \cr
              & & & -t^2& -tr&\ddots },
\ee where the translation along the diagonal is fixed by $\bra
2k-2|S | 2k \ket =-t^2$, $k\in\Z$. The real parameters $t$ and $r$
are linked by $r^2+t^2=1$ to ensure unitarity. Due to unitary
equivalence it suffices to consider $0\le t,r \le 1$. Thus $S$ is
determined by $t$. We shall sometimes write $S(t)$ to emphasize
this dependence. The spectrum of $S(t)$ is purely absolutely
continuous and is given by \be \sigma(S(t))=\Sigma(t)=\{e^{\pm i
\arccos (1-t^2(1+\cos(y))) }, y\in \T\}. \ee For this and other
properties of $S$, relations between $U_\omega$ and the physical
model alluded to in Section 1 or links with orthogonal
polynomials, see \cite{j} and Section \ref{secOPUC}. Note that the band structure (\ref{s0})
is the simplest one a unitary operator can take without being
trivial from the point of view of its spectrum,  \cite{bhj}.\\

\noindent {\bf Remarks:} \\
\noindent i) In this definition, $S$ plays the role of the free
Laplacian in the self-adjoint Anderson model. The band structure of $S$ is
inherited by $U_\omega$.\\
ii) As $t \ra 0$, $S(t)$ tends to the identity operator, whereas
as $t\ra 1$, $S(t)$ tends to a direct sum of shift operators.
Accordingly, the spectrum of $U_\omega$ is pure point if $t=0$ and
purely absolutely continuous if $t=1$, \cite{bhj}. \vskip.3cm

The localization result of \cite{j2}, when restricted to the
present one dimensional setting, says the following. If the
i.i.d.\ phases $\theta_k$ have an absolutely continuous
distribution with bounded density, i.e.\
$d\nu(\theta)=\tau(\theta)d\theta$ on $\T$, where $0\leq
\tau(\theta)\in L^\infty(\T)$, then there exists $t_0>0$ such that
$t< t_0$ implies that $\sigma (U_\omega)$ is pure point almost
surely. Thus $t$ takes the role of a disorder parameter, small $t$
corresponding to large disorder.

In the one-dimensional context, one expects to have localization
as soon as $t<1$ and $\nu$ has an absolutely continuous component.
Indeed, we show the
\begin{thm} \label{loc}
Let $U_\omega$ be defined by (\ref{beta}), (\ref{1d}) and
(\ref{s0}). If the distribution $d\nu$ of the i.i.d.\ phases
possesses a non-trivial absolutely continuous component and
supp$\,\nu$ has non-empty interior, then $U_\omega$ is pure point
almost surely, with exponentially decaying eigenfunctions.
\end{thm}

Here supp$\nu$ refers to the topological support of $\nu$, i.e.\
the set of all $\lambda$ such that $\nu(\lambda-\varepsilon,
\lambda+\varepsilon)>0$ for all $\varepsilon>0$. The requirement
that the supp$\,\nu$ contains an open set stems from the method
used here to prove positivity of Lyapunov exponents, see
Proposition~\ref{poslya} and its proof in Section~\ref{proly}
below. This assumption can be dropped, as follows from the work in
preparation \cite{Hamza/Stolz}. For example, it will be shown
there that the Lyapunov exponents are strictly positive if the
support of $\nu$ contains at least three points.

We need that $\nu$ has an a.c.\ component due to the use of a
spectral averaging argument in Section~\ref{locproof} below. One
may expect that Theorem~\ref{loc} carries over to singularly
distributed $\nu$, as was proven for discrete and continuous
one-dimensional Anderson models in \cite{CKM} and \cite{DSS},
respectively. But this would require to develop a completely
different and much more involved approach.

We further show in Section~\ref{sechalf} that our main result is
also true for unitary matrices with a similar structure defined on
$l^2(\N)$, see Theorem~\ref{loc+}. This extension allows for an
application of our analysis to random OPUC which is described in
Section~\ref{secOPUC}.

\section{Properties of the Model}

In order to prove Theorem~\ref{loc}, we need to collect some facts
about the unitary operator $U_\omega$ that are proven in
\cite{bhj} and \cite{j}.

Our definition (\ref{1d}) leads to ergodicity of the unitary
operator $U_\omega$. Indeed, introducing the shift operator $W$ on
$\Omega$ by \be\label{shift}
  W(\omega)_k=\omega_{k+2}, \ \ \ k\in\Z,
\ee we get an ergodic set $\{W^j\}_{j\in\Z}$ of translations. With
the unitary operators $V_j$ defined on the canonical basis of
$l^2(\Z)$ by \be V_j | k \ket = |k-2j\ket, \forall k\in\Z, \ee we
observe that for any $j\in\Z$ \be\label{ero}
U_{W^j\omega}=V_jU_{\omega}V_j^*. \ee Therefore, our random
operator $U_{\omega}$ is a an ergodic unitary operator. The
general theory of ergodic operators, as for example presented in
\cite{cl}, chapter V, for the self-adjoint case, carries over to
the unitary setting. In particular, it follows that the spectrum
of $U_\omega$ is almost surely deterministic, i.e.\ there is a
subset $\Sigma$ of the unit circle such that $\sigma(U_\omega) =
\Sigma$ for almost every $\omega$. The same holds for the
absolutely continuous, singular continuous and pure point parts of
the spectrum: There are $\Sigma_{ac}$, $\Sigma_{sc}$ and
$\Sigma_{pp}$ such that almost surely $\sigma_{ac}(U_{\omega}) =
\Sigma_{ac}$, $\sigma_{sc}(U_\omega) = \Sigma_{sc}$ and
$\sigma_{pp}(U_\omega) = \Sigma_{pp}$. Moreover, as shown in
\cite{j}, we can characterize $\Sigma$ in terms of the support of
$d\nu$ and of the spectrum $\Sigma(t)$ of $S(t)$.
\begin{thm} Under the above hypotheses,
the almost sure spectrum of $U_{\omega}$ consists in the set \be
\Sigma =\exp(i\,\mbox{\em supp} \ \nu)\Sigma(t)=
\{e^{i\alpha}\Sigma(t)\,\,| \,\,\alpha\in\mbox{\em supp}\ \nu \}.
\ee
\end{thm}

Let us proceed by recalling some facts concerning the generalized eigenvectors
and the associated Lyapunov exponent.
Following  \cite{bb}, \cite{bhj} and \cite{j}, we study the generalized
eigenvectors of $U_\omega$ by means of $2\times 2$ transfer matrices.  This
is possible due to the band structure of $U_\omega$.
Our unitary operators differ from those considered in the works above by
a unitary transform, so that the formulas differ a little.

Consider
\bea\label{eveq}
&&U_{\omega}\psi=e^{i\alpha}\psi, \nonumber \\
&&\psi=\sum_{k\in\Z}c_k |k\ket , \,\, c_k\in \C, \,\, \alpha\in\C.
\eea This equation is  equivalent to the relations  for all
$k\in\Z$, \be\label{trans} \pmatrix{c_{2(k+1)-1} \cr c_{2(k+1)}}=
T(\theta_{2k}^{\omega}(\alpha),\theta_{2k+1}^{\omega}(\alpha))
\pmatrix{c_{2k-1} \cr c_{2k}}, \ee where the function $T:\T^2\ra
M_2(\C)$  is defined by \bea\label{tren}
T(\theta, \eta)_{11}&=& -e^{-i\eta} \\
T(\theta, \eta)_{12}&=&\frac{r}{t}\left(e^{i(\theta-\eta)}
-e^{-i\eta}\right)\nonumber\\
T(\theta, \eta)_{21}&=&\frac{r}{t}\left(1-e^{-i\eta}\right)\nonumber\\
T(\theta, \eta)_{22}&=&-\frac{1}{t^2}\,e^{i\theta}
+\frac{r^2}{t^2}\left((e^{i(\theta-\eta)}+1)-e^{-i\eta}\right),\nonumber
\eea
and  the phases  by
\be
\theta_k^{\omega}(\alpha)=\theta_k^{\omega}+\alpha.
\ee
Note that  $ \det T(\theta_{2k}^{\omega}(\alpha),\theta_{2k+1}^{\omega}(\alpha))=
e^{i(\theta_{2k}^\omega-\theta_{2k+1}^\omega)}$ is independent of
$\alpha$ and of modulus one.

Introducing the notation
\be\label{mapt}
T(k,\omega) \equiv T(\theta_{2k}^{\omega}(\alpha),\theta_{2k+1}^{\omega}(\alpha)),
\ee
we compute for any $k\in\N$, assuming $(c_{-1}, c_0)$ known,
\bea\label{cocycle}
\pmatrix{c_{2k-1} \cr c_{2k}}&=&T(k-1,\omega)\cdots T(1,\omega)
T(0,\omega)\pmatrix{c_{-1} \cr c_{0}}
 \equiv \Phi(k, \omega)\pmatrix{c_{-1} \cr c_{0}} \\
\pmatrix{c_{-2k+1} \cr c_{-2k+2}}&=&T(-k,\omega)^{-1}\cdots
T(-2,\omega)^{-1}T(-1,\omega)^{-1}\pmatrix{c_{-1} \cr c_{0}}
\equiv \Phi(-k,\omega)\pmatrix{c_{-1} \cr c_{0}},\nonumber
\eea
with $\Phi(0, \omega)=\un$.
The dynamical system defined that way is ergodic,
\bea
&&\Phi(k,\omega)=T(0,W^{k-1}(\omega))\cdots T(0,W(\omega))T(0,\omega)
\nonumber\\
&&\Phi(-k,\omega)=T(-1,W^{-k+1}(\omega))\cdots
T(-1,W^{-1}(\omega))T(-1,\omega) \eea and the determinant of the
transfer matrices is of modulus one. As shown in \cite{bhj}, it
follows  that for any $\alpha\in\C$, the Lyapunov exponent
\be\label{lyapu} \gamma_{\omega}(e^{i\alpha}) =
\lim_{k\ra\pm\infty}\frac{1}{|k|}\ln\|\Phi(k,\omega)\| \ee almost
surely exists, has the same value for $k\to\infty$ and $k\to
-\infty$, and takes the deterministic value \be\label{lyapu2}
\gamma(e^{i\alpha}) = \lim_{k\ra\pm\infty}\frac{1}{|k|} \E
\left(\ln\|\Phi(k,\omega)\| \right). \ee

 We can
allow  spectral parameters of the form
$e^{i\alpha}=z\in\C\setminus\{0\}$, and get from classical
arguments, see {\it e.g.} \cite{cfks}, that $\gamma$ is a
subharmonic function of $z$. For other properties of
$\gamma(e^{i\alpha})$ and, in particular for its relations with
the density of states, see \cite{j}.

The link between the behaviour at infinity of the generalized
eigenvectors and the spectrum of $U_\omega$ is provided by
Sh'nol's Theorem. This is a deterministic fact which, as proven in
\cite{bhj}, carries over the the unitary operators considered here
(here $E_{\omega}(\cdot)$ is the spectral resolution of
$U_{\omega}$, which we consider to be supported on $\T$):
\begin{thm}\label{simonrevu} $\sigma(U_\omega)$ is the closure of the set
\be S_{\omega} = \{ \alpha \in \T ; U_\omega\phi = e^{i\alpha}
\phi \mbox { has a non-trivial polynomially bounded solution} \}
\ee and $E_\omega(\T \setminus S_{\omega}) =0$.
\end{thm}
A version of the Ishii-Pastur theorem suited to unitary
matrices with a band structure follows as a corollary to these arguments.
\begin{thm}\label{IP} Let $U_\omega$ be defined by (\ref{1d}), (\ref{beta})
and (\ref{s0}) and $\gamma (e^{i\alpha})$ by (\ref{lyapu}). Then
\be \Sigma_{ac} \subseteq \overline{ \{ e^{i \alpha} \in S^1 ;
\gamma (e^{i\alpha})=0 \} }^{\mbox{ess}} \enspace . \ee
\end{thm}

\vspace{.3cm}

The positivity of the Lyapunov exponent is assessed in \cite{bhj}
by means of F\"urstenberg's Theorem. The situation considered in
\cite{bhj} actually concerns  phases which are uniformly
distributed on $\T$ and this is therefore easily obtained. In the
present case, the support of $d\nu$ is more arbitrary and a more
detailed investigation is necessary. The argument leading to the
following result is presented in Section \ref{proly}.
\begin{prop}\label{poslya} Assume the supp$\,\nu$ has non-empty
interior. Then, the Lyapunov exponent $\gamma(e^{i\alpha})$ associated
wich the ergodic linear dynamical system (\ref{cocycle}) is
strictly positive for any $\alpha\in\T$.
\end{prop}
As a direct corollary, we get that \be \label{acempty}
\Sigma_{ac}=\emptyset. \ee

\section{Proof of Theorem \ref{loc}:} \label{locproof}

\setcounter{equation}{0}

We adapt an argument which, in various forms, has been used
extensively in localization proofs for various types of
one-dimensional random Schr\"odinger operators. It combines
positivity of the Lyapunov exponent with polynomial boundedness of
generalized eigenfunctions (Theorem~\ref{simonrevu}) and spectral
averaging. While implicit in the literature even earlier, this
strategy was first explicitly spelled out in \cite{sw}.

Let $\overline{\Omega}= \T^{\Z \setminus \{-1,0\}}$,
$\overline{\P} = \otimes_{k\in \Z\setminus \{-1,0\}} \nu$ and
$\overline{\omega} = (\ldots, \overline{\omega}_{-3},
\overline{\omega}_{-2}, \overline{\omega}_1, \overline{\omega}_2,
\ldots)$. We will write $(\overline{\omega}, \theta_{-1},
\theta_0)$ for $(\ldots, \overline{\omega}_{-3},
\overline{\omega}_{-2}, \theta_{-1}, \theta_0,
\overline{\omega}_1, \overline{\omega}_2, \ldots)$.

By construction $\gamma_{(\overline{\omega}, \theta_{-1},
\theta_0)}(e^{i\alpha})$ is independent of $(\theta_{-1},
\theta_0)$. It follows from Proposition \ref{poslya} that for any
$\alpha\in\T$, there exists $\overline{\Omega}(\alpha)\subset
\overline{\Omega}$ with
$\overline{\P}(\overline{\Omega}(\alpha))=1$ such that \be
\label{poslap} \gamma_{(\overline{\omega}, \theta_{-1},
\theta_0)}(e^{i\alpha}) = \gamma(e^{i\alpha})>0 \ee for all
$(\theta_{-1}, \theta_0)$ and all $\overline{\omega} \in
\overline{\Omega}(\alpha)$. Hence, by Fubini applied to
$\overline{\P} \times |\cdot|$, we get the existence of
$\overline{\Omega}_0 \in \overline{\Omega}$ with
$\overline{\P}(\overline{\Omega}_0)=1$ such that for every
$\overline{\omega} \in \overline{\Omega}_0$ there is
$A_{\overline{\omega}} \in \T$ with $|A_{\overline{\omega}}| =0$
and \be \label{poslap2} \gamma_{(\overline{\omega}, \theta_{-1},
\theta_0)}(e^{i\alpha}) > 0 \ \ \ \mbox{for all $(\theta_{-1},
\theta_0)$ and all $\alpha \in {A_{\overline{\omega}}}^C$}. \ee
Here $|\cdot|$ denotes Lebesgue-measure on $\T$.

Let us show that ${A_{\overline{\omega}}}^C$ is a support of the
spectral resolution of $U_{(\overline{\omega}, \theta_{-1},
\theta_0)}$ for Lebesgue almost every $(\theta_{-1},\theta_0)$.

We introduce the spectral measures $\mu_{\omega}^j$ associated
with $U_\omega=\int_{\T}e^{i\alpha}\ d E_\omega(\alpha)$ defined
for all $j\in\Z$ and all Borel sets $\Delta\in\T$ by \be
\mu_{\omega}^{j}(\Delta)=\bra j | E_{\omega}(\Delta) | j\ket. \ee
By construction, the variation of a random phase at one site is
described by a rank one perturbation. More precisely, dropping the
subscript $\omega$ temporarily, we define $\hat{D}$ by taking
$\theta_0=0$ in the definition of $D$: \be\label{dhat}
\hat{D}=e^{i\theta_0|0\ket\bra 0|}D=D +|0\ket\bra
0|(1-e^{-i\theta_0}), \ee so that, with the obvious notations,
\be\label{uhat} \hat{U}=\hat{D}S=e^{i\theta_0|0\ket\bra 0|}U. \ee
As for rank one perturbations of self-adjoint operators, a
spectral averaging formula holds in the unitary case. In
particular, see \cite{c} and \cite{b}, for any $f\in L^1(\T)$, \be
\int_\T \ d\theta_0 \int_\T f(\alpha)
d\mu_{(\overline{\theta},\theta_{-1}, \theta_0)}^{0}(\alpha)
=\int_\T f(\alpha)\ d\alpha. \ee By applying this to the
characteristic function of $A_{\overline{\omega}}$ we get \be 0 =
|A_{\overline{\omega}}| = \int_\T \mu_{(\overline{\omega},
\theta_{-1}, \theta_0)}^0(A_{\overline{\omega}})\, d\theta_0, \ee
implying that $\mu_{(\overline{\omega}, \theta_{-1},
\theta_0)}^0(A_{\overline{\omega}}) =0$ for every $\theta_{-1}$
and Lebesgue-a.e.\ $\theta_0$. Similarly, we get
$\mu_{(\overline{\omega}, \theta_{-1},
\theta_0)}^{-1}(A_{\overline{\omega}})=0$ for every $\theta_0$ and
Lebesgue-a.e.\ $\theta_{-1}$.

Therefore, for all $\overline{\omega}\in \overline{\Omega}_0$,
there exists $J_{\overline{\omega}}\subset \T^2$ such that
$|{J_{\overline{\omega}}}^C|=0$ and \be \label{final}
(\theta_{-1}, \theta_0)\in J_{\overline{\omega}} \Rightarrow
\mu_{(\overline{\omega},\theta_{-1},
\theta_0)}^{j}(A_{\overline{\omega}})=0, \ \ j\in \{-1,0\}. \ee

Fix $\overline{\omega} \in \overline{\Omega}_0$ and $(\theta_{-1},
\theta_0) \in J_{\overline{\omega}}$ and consider $\omega =
(\overline{\omega}, \theta_{-1}, \theta_0)$. Below we prove
\begin{lem}\label{cyc}
The subspace {\em Span} $\{|-1\ket, |0\ket\}$ is cyclic for $U_\omega$.
\end{lem}

Therefore, we deduce from (\ref{final}) that
$E_{\omega}(A_{\overline{\omega}}) = 0$. If $S_{\omega}$ is the
set from Sh'nol's Theorem~\ref{simonrevu}, we conclude that
$S_{\omega} \cap {A_{\overline{\omega}}}^C$ is a support for
$E_{\omega}(\cdot)$.

Let $\alpha \in S_{\omega} \cap {A_{\overline{\omega}}}^C$. By
Theorem~\ref{simonrevu}, $U_\omega\psi=e^{i\alpha}\psi$ has a
non-trivial polynomially bounded solution $\psi$. On the other
hand, by (\ref{poslap2}), $\gamma_{\omega}(e^{i\alpha})>0$. Thus,
by Osceledec's Theorem, every solution which is polynomially
bounded at $+\infty$ necessarily has to decay exponentially, and
the same holds at $-\infty$. Thus $\psi$ decays exponentially at
$+\infty$ and $-\infty$, and therefore is in $l^2(\Z)$ and an
eigenfunction of $U_{\omega}$. We have shown that every $\alpha
\in S_{\omega} \cap {A_{\overline{\omega}}}^C$ is an eigenvalue of
$U_{\omega}$. As $l^2(\Z)$ is separable, it follows that
$S_{\omega} \cap {A_{\overline{\omega}}}^C$ is countable.
Therefore $E_{\omega}(\cdot)$ has countable support and thus
$U_{\omega}$ is pure point spectrum, in particular \be
\label{nosing} \sigma_{sc}(U_{\omega}) = \emptyset \ \ \ \mbox{for
every $\omega \in \Omega_0 := \{ (\overline{\omega}, \theta_{-1},
\theta_0):\, \overline{\omega} \in \overline{\Omega}_0, \,
(\theta_{-1}, \theta_0) \in J_{\overline{\omega}} \} $}. \ee

From $|{J_{\overline{\omega}}}^C| =0$ and the non-triviality of
the a.c.\ component of $\nu$ we have \be \label{posmeasure} (\nu
\times \nu)(J_{\overline{\omega}}) \ge (\nu_{ac} \times
\nu_{ac})(J_{\overline{\omega}}) = (\nu_{ac} \times \nu_{ac})
(\T^2)>0. \ee

As $\overline{\P}(\overline{\Omega}_0)=1$, we conclude from
(\ref{nosing}) and (\ref{posmeasure}) that \be
\P(\sigma_{sc}(U_{\omega}) = \emptyset) \ge \P(\Omega_0) =
\int_{\overline{\Omega}_0} d\overline{\P}(\overline{\omega}) (\nu
\times \nu) (J_{\overline{\omega}}) >0. \ee

By the discussion in Section~\ref{mod} we know that spectral types
are almost surely deterministic, thus $\Sigma_{sc}= \emptyset$. We
already know $\Sigma_{ac}=\emptyset$ from (\ref{acempty}). This
proves that $U_{\omega}$ has almost surely pure point spectrum.

We still need to show that almost surely all eigenfunctions decay
exponentially. To this end, note that we actually have shown above
that the event ``all eigenvectors of $U_{\omega}$ decay at the
rate of the Lyapunov exponent'' has positive probability (as this
is true for all $\omega\in \Omega_0$). For the case of ergodic
one-dimensional Schr\"odinger operators Kotani and Simon show in
Theorem~A.1 of \cite{KoSi} that this event has probability $1$ or
$0$. In fact, only measurability needs to be shown as the event is
invariant under the ergodic transformation $W$. The proof of this
fact provided in \cite{KoSi} carries over to our model. Let us
only note that, due to Lemma~\ref{cyc}, we may use $\rho_{\omega}
= \mu_{\omega}^{-1} + \mu_{\omega}^0$ as spectral measures in
their argument. This completes the proof of Theorem~\ref{loc} up
to the

\vspace{.3cm}

\noindent {\bf Proof of Lemma \ref{cyc}:} We drop the
sub(super)scripts $\omega$ in this proof. We have to show that any
vector $|k\ket$, $k\in\Z$ can be written as a linear combination
of the vectors $U^n |j\ket$, $n\in\Z, j=-1$ and $j=0$. We compute
from (\ref{1d}) and its adjoint \bea &&U
|-1\ket=e^{-i\theta_{-2}}rt |-2\ket +e^{-i\theta_{-1}}r^2 |-1\ket+
e^{-i\theta_{0}}rt |0\ket-e^{-i\theta_{1}}t^2 |1\ket\\
&&U |0\ket=-e^{-i\theta_{-2}}t^2 |-2\ket -e^{-i\theta_{-1}}rt |-1\ket+
e^{-i\theta_{0}}r^2 |0\ket-e^{-i\theta_{1}}rt |1\ket\\\label{88}
&&U^{-1} |0\ket=e^{i\theta_{0}}(rt |-1\ket +r^2 |0\ket+
rt |1\ket-t^2 |2\ket)\\\label{77}
&&U^{-1} |-1\ket=e^{i\theta_{1}}(-t^2|-3\ket -rt |-2\ket+
r^2 |-1\ket-rt |2\ket).
\eea
Hence, using $t^2+r^2=1$,
\bea\label{up}
 &&|1\ket=\frac{e^{i\theta_{1}}}{t}\left( e^{-i\theta_{0}}r |0\ket
-(tU|-1\ket+rU |0\ket)\right)\\\label{upp}
&&|-2\ket= \frac{e^{i\theta_{-2}}}{t}\left(rU|-1\ket-tU |0\ket-
e^{-i\theta_{-1}}r |-1\ket\right).
\eea
Therefore, using (\ref{up}) in (\ref{88}) , suitable
linear combinations of $|-1\ket, |0\ket,U |-1\ket, U|0\ket $
and $U^{-1} |0\ket$ yield $|2\ket$. Similarly, $|-3\ket$ can
be obtained as a linear combination of $|-1\ket, |0\ket, U |-1\ket, U|0\ket$
and $U^{-1}|-1\ket$ using (\ref{upp}) in (\ref{77}). These manipulations
lead us from the indices $(-1,0)$ to the set $(1,2)$ in one direction
and $(-3,-2)$ in the other direction. Due to the shape of $U$, we can
iterate the process to reach any vector.\ep

\section{Half-lattice operators} \label{sechalf}

In this section we indicate how to adapt the results above to
random unitary matrices similar to (\ref{1d}), but defined on
$l^2(\N_0)$, in the same spirit as in \cite{bhj}.

Let $S^+$ be the unitary defined on $l^2(\N_0)$, $\N_0 =
\{0,1,2,\ldots\}$, in the canonical basis $\{|j\ket\}_{j\in\N_0}$
by the matrix \be\label{s+} S^+=\pmatrix{  -r & rt&-t^2 & & &\cr
               t& r^2  &-rt & & &\cr
                & rt   & r^2 & rt& -t^2&\cr
                &-t^2  & -rt& r^2&-rt& \cr
                &      &    &tr &r^2& \cr
                &      &    & -t^2  &-rt&\ddots },
\ee where the dots mean repetition of the last $4\times 2$ block,
as in (\ref{s0}). Then one considers \be\label{u+}
U_\omega^+=D^+_{\omega}S^+\ \ \ \mbox{with} \ \ \
D^+_{\omega}=\mbox{ diag }\{e^{-i\theta_k^\omega}\}, \ee where the
random phases $\theta_k^\omega$ are given by (\ref{beta}), for
$k\in \N_0$. The generalized eigenvectors $\psi=\sum_{k\geq 0}c_k\
|k\ket$
 defined by
\be\label{adj} U_\omega^+\psi=e^{i\alpha}\psi \ee give rise to the
same dynamical system (\ref{cocycle}) on the coefficients $c_k$.
Starting from $(c_1, c_2)^T$, we have \be\label{re}
\pmatrix{c_{2(k+1)-1} \cr c_{2(k+1)}}=
T(\theta_{2k}^{\omega}(\alpha),\theta_{2k+1}^{\omega}(\alpha))
\pmatrix{c_{2k-1} \cr c_{2k}}, \quad k=1,2,\ldots \ee
where the
transfer matrix $T(\theta,\eta)$ is given by (\ref{tren}) and
 $\theta_k^\omega(\alpha)=\theta_k^\omega + \alpha$ as before. This relation
must supplemented by
\be
\pmatrix{c_1\cr c_2}=c_0\pmatrix{(e^{-i(\theta_1+\alpha)}+
re^{-i(\theta_1-\theta_0)})/t\cr
(e^{-i(\theta_1+\alpha)}+re^{-i(\theta_1-\theta_0)})r/t^2-
(r+e^{i(\alpha+\theta_0})/t^2},
\ee
where $c_0$ is free, because of the boundary condition at component $0$.
The (forward) Lyapunov
exponent $\gamma(e^{i\alpha})$ corresponding to (\ref{re}) defined by
(\ref{lyapu}) exists
almost surely and the conclusions of Theorems \ref{simonrevu} and \ref{IP}
remain true for $U^+_\omega$.

The first difference/simplification with respect to the operator
$U_\omega$ defined on
the whole lattice is that $U^+_\omega$ admits a cyclic vector
\begin{lem}
The vector $|0\ket $ is cyclic for $U_\omega^+$.
\end{lem}
{\bf Proof: } One first checks that $ |1\ket
=\frac{e^{i\theta_1}}{t}\left(U^+_\omega|0\ket+re^{-i\theta_0}|0\ket
\right), $ and then one concludes as in the proof of Lemma
\ref{cyc}. \ep

As a consequence, we get

\begin{thm} \label{loc+}
Let $U_\omega^+$ be defined by (\ref{s+}), (\ref{u+}) and
(\ref{beta}). If the distribution $d\nu$ of the i.i.d. phases
possesses a non-trivial absolutely continuous component and its
support has non-empty interior, then $U_\omega^+$ is pure point
almost surely, with exponentially decaying eigenfunctions.
\end{thm}
{\bf Proof:} The proof is virtually the same as that of Theorem
\ref{loc}. Due to cyclicity of $|0\ket$ it suffices to average
over the single phase $\theta_0$, which leads to some
simplifications. \ep

\section{Application to OPUC} \label{secOPUC}

The previous extension of our result to $l^2(\N_0)$ was aimed to
pave way for the applications of our localization results to
orthogonal polynomials on the unit circle (OPUC) with respect to
an infinitely supported probability measure $d\mu$. Such
polynomials $\Phi_k$ are determined via the Szego recursion
$\Phi_{k+1}(z) = z\Phi_k(z)-\overline{\alpha}_k \Phi_k^*(z)$,
$\Phi_0=1$, by a sequence of complex valued coefficients
$\{\alpha_k\}_{k\in\N_0}$, such that $|\alpha_k|< 1$, called
Verblunski coefficients, which also characterize the measure
$d\mu$, see \cite{s}. This latter relation is encoded in a five
diagonal unitary matrix $C$ on $l^2(\N_0)$ representing
multiplication by $z\in S^1$: The measure $d\mu$ arises as the
spectral measure $\mu(\Delta)= \bra 0|E(\Delta)|0\ket$ of the
cyclic vector $|0\ket$ of $C$. This matrix is the equivalent of
the Jacobi matrix in the case of orthogonal polynomials with
respect to a measure on the real axis, and it is called the CMV
matrix, after \cite{cmv}.

In case the Verblunski coefficients all have the same modulus ,
i.e. \be \alpha_k=re^{i\eta_k}, \ \ \ k=0,1, \ldots \ee the
corresponding CMV matrix reads \be C=\pmatrix{  re^{-i\eta_0} &
rte^{-i\eta_1} & t^2 & & & \cr
               t& -r^2e^{i(\eta_0-\eta_1)}  &-rte^{i\eta_0} & & &\cr
                & rte^{-i\eta_2} & -r^2e^{i(\eta_1-\eta_2)} & rte^{-i\eta_3}& t^2 &\cr
                &t^2 & -rte^{i\eta_1}& -r^2e^{i(\eta_2-\eta_3)}&-rte^{i\eta_2}& \cr
               & & &rte^{-i\eta_4} & -r^2e^{i(\eta_3-\eta_4)} & \cr
               & & &t^2 &-rte^{i\eta_3} &\ddots}.
\ee Now, changing from the canonical basis $\{|j\ket\}_{j\in\N_0}$
to $\{e^{i\beta_j}|j\ket\}_{j\in\N_0}$ by means of the unitary $B$
defined by $B|j\ket=e^{i\beta_j}|j\ket$, $j=0,1,\cdots $, we get
\be\label{bcb} B^{-1}CB= \pmatrix{  re^{-i\eta_0} &
rte^{-i\eta_1}e^{i(\beta_1-\beta_0)} &
                                                    t^2e^{i(\beta_2-\beta_0)}  & \cr
               te^{i(\beta_0-\beta_1)}& -r^2e^{i(\eta_0-\eta_1)}
                                         &-rte^{i\eta_0}e^{i(\beta_2-\beta_1)}  &\cr
                & rte^{-i\eta_2}e^{i(\beta_1-\beta_2)} & -r^2e^{i(\eta_1-\eta_2)} &
                  & \cr
                &t^2e^{i(\beta_1-\beta_3)} & -rte^{i\eta_1}e^{i(\beta_2-\beta_3)}&
           \ddots& }.
\ee Then, by choosing the $\beta_j$'s suitably, the matrix
(\ref{bcb}) becomes the negative of a matrix of the form
(\ref{u+}): \be \label{-u+} -U^+=\pmatrix{ re^{-i\theta_0} &
-re^{-i\theta_0}t&t^2e^{-i\theta_0} & & & \cr
                  -te^{-i\theta_1}& -r^2e^{-i\theta_1}  &rte^{-i\theta_1} & & &\cr
                   & -rte^{-i\theta_2} & -r^2e^{-i\theta_2} &
                             -rte^{-i\theta_2}& t^2e^{-i\theta_2} &\cr
                   & t^2e^{-i\theta_3} & rte^{-i\theta_3}& -r^2e^{-i\theta_3}
                                             & rte^{-i\theta_3} &\cr
                   & & &- tre^{-i\theta_4} & -r^2e^{-i\theta_4}&\cr
                   & & &t^2e^{-i\theta_5} &rte^{-i\theta_5}&\ddots }.
\ee Here, as seen from the diagonal elements, the phases
$\theta_k$ are given in terms of the phases of the Verblunski
coefficients by (set $\eta_{-1}=0$) \be\label{theta}
\theta_k=\eta_k-\eta_{k-1}, \ \ k=0,1,2, \cdots , \ee or,
equivalently \be \eta_k=\theta_{k}+\theta_{k-1}+\cdots +\theta_0,
\ \ k=0,1,2, \cdots \ee The terms in $t^2$ require
\bea\label{beta1}
& & \beta_1-\beta_0=\theta_1+\pi\nonumber\\
& & \beta_{2k+1}-\beta_{2k-1}=\theta_{2k+1},\ \ k=1,2, \ldots ,\nonumber \\
& & \beta_{2k+2}-\beta_{2k}=-\theta_{2k},\ \ k=0,1, \ldots, \eea
where $\beta_0$ is free. Explicitly, for $k\geq 0$, \bea
\beta_{2k+1}&=&\theta_{2k+1}+\theta_{2k-1}+\cdots +\theta_1+\beta_0+\pi\nonumber\\
\beta_{2k+2}&=&-(\theta_{2k}+\theta_{2k-2}+\cdots +\theta_0). \eea
It is straightforward to check that (\ref{theta}) and
(\ref{beta1}) form a consistent choice in the sense that all terms
in $rt$ in (\ref{bcb}) and (\ref{-u+}) agree. Assuming the
$\theta_k$'s are i.i.d.\ random variables, Theorem \ref{loc+}
applies to this case and yields
\begin{prop}
Let ${\alpha_k(\omega)}_{k\in\N_0}$ be random Verblunski
coefficients of the form \be\alpha_k(\omega)=re^{i\eta_k(\omega)},
\ \ \  0<r<1, \ \ \ k=0,1,2,\ldots \ee whose phases are
distributed on $\T$ according to \be \eta_k(\omega) \sim d\nu *
d\nu * \cdots
* d\nu\,, \ \ \ \mbox{($k+1$ convolutions)} \ee and
$d\nu$ is a probability measure with non-trivial a.c.\ component
and such that its support has non-empty interior. Then, the random
measure $d\mu_\omega$ on $S^1$ with respect to which the
corresponding random polynomials $\Phi_{k,\omega}$ are orthogonal
is almost surely pure point.
\end{prop}
{\bf Remark:} Other localization results for random polynomials on
the unit circle, \cite{ps}, \cite{t}, \cite{su} are proven for
independent Verblunski coefficients. Moreover, the results of
\cite{su} and \cite{ps} require rotational invariance of the
distribution of the Verblunski coefficients in the unit disk. By
contrast, the proposition above holds for strongly correlated
random Verblunski coefficients.

\section{ Proof of Proposition~\ref{poslya}}\label{proly}

We want to apply F\"urstenberg's Theorem. Since the latter is
stated for real valued matrices, we proceed as in \cite{bhj} and
introduce the mapping $\tau: M_2(\C)\ra M_4(\R)$ defined by \be
\pmatrix{a & b \cr c & d }\rightarrow
  \pmatrix{\Re(a)I + \Im(a)J & \Re(b)I +
    \Im(b)J \cr \Re(c)I + \Im(c)J & \Re(d)I + \Im(d)J },
\ee where \be I = \left( \begin{array}{cc} 1 & 0 \\ 0 & 1
\end{array} \right) \enspace , \enspace J = \left(
\begin{array}{cc} 0 & 1 \\ -1 & 0 \end{array} \right)\enspace .
\ee This mapping is a homeomorphism from  $M_2(\C)$ to $\tau(
M_2(\C))$ and, in particular, a group homeomorphisms from the set
of matrices in $M_2(\C)$ with determinant of modulus one to the
set of matrices in $M_4(\R)$ with determinant of modulus one.

Let $G_{\alpha,\nu}$ be the closed group generated by the matrices
$T(\theta,\eta)$ from (\ref{tren}), where $\theta$ and $\eta$ vary
in $\alpha + \mbox{supp}\,\nu$. As discussed in \cite{bhj},
F\"urstenberg's Theorem implies that $\gamma(e^{i\alpha})>0$ if it
can be shown that $G_{\alpha,\nu}$ is strongly-irreducible and
non-compact. As supp$\,\nu$ has non-empty interior and thus
$G_{\alpha,\nu}$ has a non-trivial connected component it suffices
to show
\begin{lem}\label{suff}
The group $\tau(G_{\alpha,\nu})$ is irreducible and non-compact.
\end{lem}
To show the first statement, one uses  arguments of the same type
as those developed in \cite{bhj}. The set $\alpha +
\mbox{supp}\,\nu$ contains a non-empty open interval $\I$ and it
will suffice to only work with $\theta, \eta \in \I$.

Let us assume there exists a strict subspace $V$ of $\R^4$ which
is invariant under $\tau(T(\theta,\eta))$, for all $(\theta, \eta)
\in \I\times \I$. The mapping $\tau(T)$ is smooth in $\I\times \I$
and it follows by approximating derivatives with finite
differences that $V$ is also left invariant by
$\partial_\theta^r\partial_\eta^s \tau(T(\theta,\eta))$ for all
$r, s\in\N_0$ and for all $(\theta,\eta)\in \I\times \I$.

In particular, we compute for $\eta=\theta\in \I$, \bea
\label{eqdiag1}
\tau(T(\theta,\theta))&=&A_0+A_1\sin(\theta)+A_2\cos(\theta),\\
\label{eqdiag2}
\partial_\theta\tau(T(\theta,\eta))|_{\eta=\theta}&=&B_0+B_1\sin(\theta)+
B_2\cos(\theta),
\eea
with
\be
A_0=\pmatrix{0 &0 &r/t &0 \cr
0 &0 &0 &r/t \cr
r/t &0 &2r^2/t^2 & 0\cr
 0&r/t &0 &2r^2/t^2 }, A_1=\pmatrix{0 &1 &0& r/t  \cr
-1 &0 &-r/t&0 \cr
0 &r/t &0 & -1\cr
 -r/t&0 &1 &0 },
\ee
$A_2=-(A_0+\un)$, and
\bea
& &B_0=\pmatrix{0 &0 &0 &r/t \cr
0 &0 &-r/t&0 \cr
0 &0 &0&r^2/t^2 \cr
 0& 0 &-r^2/t^2&0 }, B_1=\pmatrix{0 &0 &0 &0 \cr
0 &0 &0&0 \cr
0 &0 &1/t^2&0 \cr
 0& 0 & 0&1/t^2 },\nonumber \\
& &B_2=\pmatrix{0 &0 &0 &0 \cr
0 &0 &0&0 \cr
0 &0 & 0&-1/t^2 \cr
 0& 0 & 1/t^2&0 }.
\eea It follows that $V$ is invariant under $A_0$, $A_1$, $A_2$,
$B_0$, $B_1$ and $B_2$ (for example by differentiating the right
hand sides of (\ref{eqdiag1}) and (\ref{eqdiag2}) two more times).
Since these matrices are real (anti) self-adjoint, they leave
$V^\perp$ invariant as well. Hence, $V$ and $V^\perp$ are
generated by real eigenvectors of these matrices, if they are
diagonalizable over $\R$. Note that $B_1$ is diagonal in the
canonical basis denoted by $\{e_j\}_{j\in{1,2,3,4}}$.

If $V$  is one-dimensional, it is generated by one vector which is
either in the subspace $ \bra e_1, e_2\ket $ or in the $ \bra e_3,
e_4 \ket $. As neither of these subspaces is invariant under
$A_1$, this is impossible. The same is true for $V^\perp$, so that
$V$ must be two-dimensional. $V$ cannot coincide with any of the
previously considered subspaces, by the same argument, so the only
possibility left is \be V= \bra w_1, w_2 \ket, \ \ \mbox{ where }\
\ w_1=\pmatrix{\alpha \cr \beta\cr 0\cr 0 }, \ w_2=\pmatrix{0 \cr
0\cr \gamma\cr \delta} \ \ \mbox{ and  }\ \ \alpha, \beta,\gamma,
\delta \in\R. \ee But, to have $A_1w_1\in V$, there must exist $a,
b\in\R$ such that $A_1w_1=aw_1+bw_2$, i.e. \be \pmatrix{\beta\cr
-\alpha\cr \beta r/t\cr -\alpha r/t}= \pmatrix{a\alpha \cr a\beta
\cr b\gamma\cr b\delta}. \ee The first two components imply
$(1+a^2)\beta=0$, thus $\beta=0$ and $\alpha=0$, which is absurd.
Hence $\tau(G_{\alpha,\nu})$ is irreducible.
\\

\vspace{.2cm} Let us now turn to the second statement of Lemma
\ref{suff}. The non-compactness of $\tau(G_{\alpha,\nu})$ and
$G_{\alpha,\nu}$ being equivalent, we can choose to work on
$G_{\alpha,\nu}$. We will actually show a much stronger statement.
For this, pick any fixed $\theta$ and $\eta$ on the torus with
$\theta\not= \eta$. Consider the subgroup $G(\theta,\eta)$ of
$GL(2,\C)$ generated by $T(\theta,\theta), T(\eta,\eta),
T(\theta,\eta)$ and $T(\eta,\theta)$, where the latter are defined
through (\ref{tren}). As $\alpha + \mbox{supp}\,\nu$ contains at
least two points, non-compactness of $G_{\alpha,\nu}$ will follow
from non-compactness of $G(\theta,\eta)$.

With the abbreviations $x:= e^{-i\theta}$ and $z:= e^{-i\eta}$ the
first of them takes the form
\[ T(\theta,\eta) = \left( \begin{array}{ll} -z & \frac{r}{t}
(\bar{x}z-z) \\ \frac{r}{t}(1-z) & \frac{r^2}{t^2}(\bar{x}z+1-z)
-\frac{1}{t^2} \bar{x} \end{array} \right), \] with $\det
T(\theta,\eta) =\bar{x}z$, and similarly for the other three
generating matrices.

Calculations (some lengthy) show that
\[ C:= T(\theta,\theta) T(\theta,\eta)^{-1} = \left(
\begin{array}{ll} x\bar{z} & 0 \\ \frac{r}{t}(x\bar{z}-1) & 1
\end{array} \right) \in G(\theta,\eta), \]
\[ E := T(\eta,\theta)^{-1} T(\theta,\theta) = \left(
\begin{array}{ll} 1 & \frac{r}{t} (1-\bar{x}z) \\ 0 & \bar{x}z
\end{array} \right) \in G(\theta,\eta), \]
\[ L := CE = \left( \begin{array}{ll} x\bar{z} & \frac{r}{t}
(x\bar{z}-1) \\ \frac{r}{t} (x\bar{z}-1) &
\bar{x}z-\frac{r^2}{t^2} |x\bar{z}-1|^2  \end{array} \right) \in
G(\theta,\eta),
\]
\[ J := EC = \left( \begin{array}{ll} x\bar{z} - \frac{r^2}{t^2}
|\bar{x}z-1|^2 & \frac{r}{t}(1-\bar{x}z) \\
\frac{r}{t}(1-\bar{x}z) & \bar{x}z \end{array} \right) \in
G(\theta,\eta). \]

Note that $\det L = \det J = 1$ and that $J^{-1} = L^*$. Thus we
get the self-adjoint element $K := J^{-1}L$ of $G(\theta,\eta)$.
In fact, $K$ is positive definite and $\det K=1$.

More calculation shows that
\begin{eqnarray*}
\mbox{tr}\,K & = & 1 + \frac{2r^2}{t^2} |x\bar{z}-1|^2 +
\left|x\bar{z}- \frac{r^2}{t^2} |x\bar{z}-1|^2\right|^2 \\
& = & 2 + \frac{r^2}{t^4} |x\bar{z}-1|^4.
\end{eqnarray*}

As $\theta \not= \eta$ and therefore $x\bar{z} \not= 1$ we
conclude that tr$\,K >2$. Positivity of $K$ implies that it has an
eigenvalue strictly bigger than $1$. Thus, containing all powers
of $K$, the group $G(\theta,\eta)$ is non-compact. \ep


\begin{thebibliography}{xxxxxxx}
\bibitem[AM]{am} Aizenman, M., Molchanov, S.: Localization at large disorder
and at extreme energies: an elementary derivation, {\it Commun.
Math. Phys.} {\bf 157}, 245-278, (1993).
\bibitem[B]{b} Bourget, O.: Singular continuous Floquet operator for periodic
Quantum systems,{\it J. Math. Anal. Appl.} {\bf 301}, 65-83,
(2005).
\bibitem[BB]{bb} Blatter, G., Browne, D.:  Zener tunneling and
    localization in small conducting rings, {\it Phys. Rev. B} {\bf 37},
 3856, (1988).
\bibitem[BHJ]{bhj} Bourget, O., Howland, J.S., Joye, A.: Spectral Analysis of Unitary
Band Matrices, {\it Commun. Math. Phys.} {\bf 234}, 191-227
(2003).
 .
\bibitem[C]{c} Combescure, M.: Spectral Properties of a Periodically Kicked
Quantum Hamiltonian, {\it J. Stat. Phys.} {\bf 59}, 679-690,
(1990).
\bibitem[CFKS]{cfks} Cycon, H.L., Froese, R.G., Kirsch, W., Simon,
B.:
    {\it Schr\"odinger Operators}, Springer Verlag, 1987.
\bibitem[CKM]{CKM} Carmona, R., Klein A., Martinelli, F.: Anderson
localization for Bernoulli and other singular potentials, {\it
Commun. Math. Phys.} {\bf 108}, 41-66, (1987).
\bibitem[CL]{cl} Carmona, R., Lacroix, J.:  {\it Spectral theory of random
    Schrodinger Operators}, Birkh\"auser, 1990.
\bibitem[CMV]{cmv} Cantero, M.J., Moral, L., Vel\'azquez, L.:
Five-Diagonal Matrices and Zeros of Orthogonal Polynomials on the
Unit Circle, {\it Linear Algebra Appl.} {\bf 362}, 29-56, (2003).
\bibitem[DSS]{DSS} Damanik, D., Sims, R., Stolz, G.: Localization of
one-dimensional, continuum, Bernoulli-Anderson models, {\it Duke
Math. J.} {\bf 114}, 59-100, (2002).
\bibitem[GT]{gt} Geronimo, J.S., Teplyaev, A.: A Difference Equation
Arising from the Trigonometric Moment Problem Having Random
Reflection Coefficients-An Operator Theoretic Approach, {\it J.
Func. Anal.} {\bf 123}, 12-45, (1994).
\bibitem[HS]{Hamza/Stolz} Hamza, E., Stolz, G.: in preparation
\bibitem[J1]{j} Joye, A.: Density of States and Thouless Formula
for Random Unitary Band Matrices, {\it Ann. Henri Poincar\'e} {\bf 5},
347--379, (2004).
\bibitem[J2]{j2} Joye, A.: Fractional Moment Estimates
for Random Unitary Band Matrices, {\it Lett. Math. Phys.}, to appear.
\bibitem[KS]{KoSi} Kotani, S., Simon, B.: Localization in general
one-dimensional random systems, {\it Commun. Math. Phys.} {\bf
112}, 103--119, (1987).
\bibitem[S1]{s} Simon, B.: Orthogonal Polynomials on the Unit Circle,
Vol. 1 and 2, {\it AMS Colloquium Series, American Mathematical
Society, Providence, RI}, to appear.
\bibitem[S2]{ps} Simon, B.: Aizenman's Theorem for Orthogonal
Polynomials on the Unit Circle, preprint, mp-arc 04-386.
\bibitem[S3]{s3} Simon, B.: OPUC on one foot, preprint, mp-arc
05-78.
\bibitem[SW]{sw} Simon, B., Wolff, T.: Singular Continuous Spectrum under
Rank One Perturbations and Localization for Random Hamiltonians, {\it Commun.
Pure Appl. Math.} {\bf 39}, 75--90, (1986).
\bibitem[Su]{su} Stoiciu, M.: The statistical distribution of the
zeros of random paraorthonormal polynomials on the unit circle,
preprint, mp-arc 04-405.
\bibitem[T]{t} Teplyaev, A. V.: the Pure Point Spectrum of Random Polynomials
orthogonal on the Circle, {\it Soviet. Math. Dokl.} {\bf 44},
407-411, (1992).

 \end{thebibliography}
\end{document}